\documentclass{iau}
\usepackage{graphicx, url} 

\title[IAUS291.~~Pulsars in Globular Clusters] %% short title %%
{The pulsar population in Globular Clusters and in the Galaxy} %% full title %%

\author[P. C. C. Freire]  %% short author list %%
{Paulo C. C. Freire}
\affiliation{Max-Planck-Institut f\"ur Radioastronomie, auf dem H\"ugel 69, 53121, Bonn, Germany  \\ email: {\tt pfreire@mpifr-bonn.mpg.de}}

\pubyear{2012}
\volume{291}  %% insert here IAU Symposium No.
\jname{\mbox{Neutron Stars and Pulsars: Challenges and Opportunities after 80 years}}
\editors{J. van Leeuwen, ed.} 
\begin{document}

\maketitle

%% -- Abstract ----------------------------------
\begin{abstract}
In this paper, I review some of the basic properties of the
pulsar population in globular clusters (GCs) and compare it with the
the Galactic disk population. The neutron stars (NSs) in
GCs were likely formed - and appear to continue forming - in
highly symmetric supernovae (SNe), likely from
accretion-induced collapse (AIC). I review the many pulsar
finds and discuss some particularly well populated GCs and
why they are so.
I then discuss some particularly interesting objects, like
millisecond pulsars (MSPs) with eccentric orbits, which
were heavily perturbed by passing stars. Some of these
systems,
like NGC~1851A and NGC~6544B, are almost certainly the
result of exchange interactions, i.e., they
are witnesses to the very same processes that created the
large population of MSPs in the first place. I also
review briefly the problem posed by the presence of young
pulsars in GCs (with a special emphasis on a sub-class of
young pulsars, the super-energetic MSPs), which suggest
continuing formation of NSs in low-velocity SNe.
In the final section, I discuss the possibility of an
analogous population in the Galaxy and highlight a particularly
interesting case, PSR J1903+0327, where
the primary neutron star appears to have formed with a small-velocity
kick and small fractional mass loss.
Systems with primary NSs formed in electron-capture SNe should
constitute a distinct low-velocity Galactic population akin in many
respects to the GC population. Current high-resolution surveys
of the Galactic plane should be able to detect it clearly.

%% add here a maximum of 10 keywords, to be taken form the file <Keywords.txt>
\end{abstract}

% add below any authors, subjects and objects for indexing 
%   add more lines if necessary
%   but leave all lines commented out
%\index[author]{LastName1, Initials|textbf}
%\index[author]{LastName2, Initials|textbf}
%\index[subject]{Keyword1}
%\index[subject]{Keyword2}
%\index[object]{Object1}
%\index[object]{Object2}

\firstsection % if your document starts with a section,
              % remove some space above using this command.
\section{The Pulsar population in Globular Clusters}

Globular clusters (GCs) are spherical, bound swarms of
stars containing from $10^4$ to $\sim 5 \times 10^6$ stars.
Near their centers the star density is normally over $10^3$
(and in some cases $10^6$!) per cubic parsec.
They orbit the centers of most Galaxies through the
Universe, about 200 orbit our Milky Way (\cite{har96}).
Of these, 28 clusters contain a total of 144
known radio pulsars\footnote{Our updated
reference list is at
{\url{http://www.naic.edu/~pfreire/GCpsr.html}}}.

The GC pulsar population differs from the Galactic
disk population in two main ways:
\begin{enumerate}
\item It is {\em much} older than the Galactic population. This is
to be expected given the great age of the stellar population in
GCs. This population is so old that, with a few important
exceptions (discussed below), only recycled pulsars,
which have lifetimes of many Gyr, are still detectable as radio pulsars.
\item It is a very {\em abundant} population. Per unit mass there
appear to be two to three orders of magnitude
more pulsars in GCs as in the Galactic disk. This also applies
to X-ray sources, the progenitors of MSPs.
\end{enumerate}

The reasons for this latter fact, and its many consequences,
are discussed in detail below.

\section{Large neutron star population in globular clusters}

\subsection{Exchange encounters}

In the early 1970's, the Uhuru and OSO-7 X-ray satellites revealed the
presence of several X-ray sources in GCs (e.g.~\cite{gmg+74}).
One of the co-authors (H. Gursky) recognized in 1973 that,
compared to the stellar mass in the Galaxy, this represented
a large overabundance of X-ray sources. In 1975
G. Clark\nocite{cla75} suggested that, given the extremely
high stellar densities in the cores of some GCs, it
occasionally happens that many old, dead NSs lurking in the core
``collide" with a binary, disrupt it and acquire a (new) companion.
The latter then evolves, fills its Roche lobe and
starts transferring matter to the NS, forming a low-mass X-ray
binary (LMXB).

\subsection{Origin of large NS population}

An important question already considered in these early
studies is the origin of all these lurking NSs.
In the early 1970's some pulsar proper motions had already
been measured (e.g.,~\cite{mtv74}) and these already hinted at
the fact that many NSs form with kick velocities of hundreds
of km per second. This has since been confirmed by many
subsequent studies, with ever-increasing samples and better
quality measurements (\cite{ll94,hllk05}).
Such objects would not be retained in GCs, which have escape
velocities of a few tens of km\,s$^{-1}$ --- unless they are
anchored by massive companions (\cite{dh98}). Even with
such anchoring, there seems to be a large excess of NSs
(\cite{prp02}), which lead to a suggestion, still valid
(originally by \cite{katz75}) that NSs in GCs
are forming through a low-velocity channel.
This is now thought (\cite{plp+04}) to be accretion-induced collapse
(AIC) of massive Oxygen-Neon-Magnesium WDs
that become unstable once they approach
the Chandrasekhar limit. This results in
electron capture SNe (\cite{phlh08,lan12}), not
Type 1a SNe --- there is not enough carbon
available to power thermonuclear deflagrations.
We will come back to this issue later.

\section{Millisecond pulsars - in the Galaxy and in GCs}

Soon after the discovery of the first MSP,
B1937+21 in 1982 (\cite{bkh+82}), it was suggested that
MSPs are the end stages of the
evolution of LMXBs (\cite{acrs82}). This is consistent with
the finding that, unlike in the case of normal pulsars,
most MSPs are found in binary systems. Here the mystery
is why some of these objects (including B1937+21
itself) are found to be isolated; there is no satisfactory
answer to this question yet. Furthermore, MSPs have magnetic
fields much smaller than those generally found in the
normal pulsar population; which means that accretion somehow
``buries" the magnetic field. This process is not well
understood.

If MSPs really evolve from LMXBs, then GCs,
with their LMXB over-abundance, should also contain many
MSPs. Finding them was difficult, given the great distances
to GCs. Nevertheless, if a bright radio source is discovered
in a GC, then it is likely a pulsar. In 1985, twelve nearby
GCs were imaged with the Very Large Array (VLA, \cite{hhb85}),
with one likely candidate in M28, B1821$-$24. Two years later, using the
Jodrell Bank 75-m radio telescope, the discovery
of radio pulsations at a period of 3.05 ms and DM of  120 cm$^{-3}$pc
confirmed it as the first pulsar in a GC (\cite{lbm+87}).
In the following year, PSR~B1620$-$26 is discovered in the globular
cluster M4 (\cite{lbb+88}). This pulsar, with a spin period
of 11 ms, is in a 191-day orbit with a white dwarf
(WD). Interestingly,
this binary system appears to be orbited by a Jovian-type
planet in a very wide orbit (\cite{srh+03}).

\subsection{Finding more}

The pulsars in GCs are faint owing to their
great distances. This means that we are limited first and foremost
by the sensitivity of the radio telescope and receiver
used for the survey. This can be, to some extent,
compensated by the fact that the pulsars
are located in a small region around the cluster center
(owing to the effects of mass segregation, see e.g, \cite{fcl+01})
which normally fits very well inside
a single radio beam of even the largest radio telescopes.
This allows for deep, multi-hour search observations, which
can still be made sensitive to binary pulsars using acceleration search
techniques (\cite{clf+00,rem02}). This is the reason why
for most of the last 20 years we knew more MSPs in
clusters than in the Galactic disk.

\subsection{Notable clusters and why they are so}

The early leader in pulsar discoveries
was 47~Tucanae (\cite{mlj+90,mlr+91,rlm+95,clf+00}),
which now has a total of 23 known pulsars, all of them with spin periods
shorter than 7.6 ms. Of these, 15 are in binary systems, at least
five of them eclipsing. The timing of these pulsars (\cite{fcl+01})
allowed X-ray detections of {\em all} of the MSPs (\cite{hge+05,bgh+06}),
a study of the dynamics of the cluster (\cite{fck+03}) and the
first detection of any sort of interstellar medium in a globular
cluster, after more than 60 years of searches (\cite{fkl+01}).

However, radio maps of a large group of GCs (\cite{fg00})
suggested that the heavily obscured GC
Terzan~5 had a large pulsar population, but of these
only two were known before that study
(\cite{ljm+90,lmbm00}). Sensitive observations at 2 GHz
with the GBT have since discovered 32 pulsars (\cite{rhs+05,hrs+06}).
These include Terzan~5~ad (\cite{hrs+06}): with a
spin frequency of 716 Hz; this broke the 24-year old record
set by the original MSP, B1937+21. It is thought that $\sim 100$
pulsars remain to be discovered in Terzan~5 (\cite{blc11}).

Using the same observing system 8 new pulsars were
found in NGC~6440
and NGC~6441 (\cite{frb+08}) and 11 new pulsars in M28 alone!
Freire et al. (2008a) found that, for any particular luminosity
threshold, there are nearly as many pulsars in NGC~6440 and NGC~6441
as there are in Terzan 5. The latter appears to be
exceptional because of its smaller distance to the Solar System
(5.5 kpc), as opposed to 8.2 kpc for NGC~6440 and 13.5 kpc for
NGC~6441. This results highlights the fact that these surveys
are strongly limited by sensitivity.

It is nevertheless clear that, after correcting for distance,
some GCs have many more pulsars
than the average. Although we don't know all the factors involved
in producing a large pulsar population,
the stellar encounter rate has some predictive power
not only in estimating the number of MSPs, but
of other types of objects (e.g., \cite{dav95}), particularly
X-ray binaries (e.g., \cite{pla+03}).

\subsection{Clusters with isolated pulsars}

The pulsar populations of some clusters
[NGC 7078 (\cite{and93}), NGC 6624 (\cite{bbl+94,lfrj12}),
NGC~6517 (\cite{lrfs11})
and NGC 6752 (\cite{dlm+01,dpf+02})] are dominated by isolated
pulsars, while others [like 47~Tuc (\cite{clf+00}),
M62 (\cite{pdm+03,lfrj12}), M3, M5 and
M13 (\cite{kapw91,awkp97,hrs+07})] are dominated by binaries.
The distinguishing characteristic appears to be
the core density; most of the GCs where there are
more isolated pulsars appear to be core collapsed. The high
stellar density at the core might be 
disrupting previously formed binary MSPs. Another possible
disrupter are central black hole binaries, as has been suggested in
the case of NGC~6752 (\cite{cpg02}).

Furthermore, many of the GCs dominated by
isolated pulsars appear to have a slower pulsar
population (\cite{ver03}), even in surveys where such clusters
are observed with uniform sensitivity to fast-spinning pulsars,
as in the comparison of M3/M5/M13 with M15 (\cite{hrs+07}) or
NGC 6440/6441 with Terzan~5 (\cite{frb+08}). It is likely that
the high stellar densities of some GC cores are also disrupting
X-ray binaries, leaving behind partially recycled pulsars.

\subsection{Young pulsars in globular clusters}

Some pulsars in GCs [PSR~B1718$-$19 in NGC~6342 (\cite{lbhb93}),
PSR~B1820$-$30B (\cite{bbl+94}) and J1823$-$3021C (\cite{lfrj12})
in NGC~6624 and B1745$-$20 in NGC~6440 (\cite{lmd96})] have
characteristics very similar to the normal pulsars found in the
Galactic disk: periods of the order of a few tenths of a
second, magnetic fields of the order of 10$^{11-12}$ G
and characteristic ages of a few times $10^7$~yr - about $10^3$
times younger than the stellar population in GCs. Clearly
these pulsars cannot have formed in recent
iron core collapse SNe --- there have not been
any since the first few tens of Myr of the histories of these
clusters. The partial recycling described above could be spinning
up old, dead NSs just enough to make them active
radio pulsars, but without going on for long enough to bury
their magnetic fields (\cite{lmd96}).

An alternative hypothesis, also discussed in 
Lyne et al. (1996) is ongoing formation of new NSs
through e-capture SNe. If this hypothesis is correct,
then the SNe must have small kicks, otherwise, the formation
rates required by the observed population would be unrealistic
(\cite{blt+11}).

\section{Exotic pulsars in GCs}

\subsection{Eccentric binary MSPs \& their uses}

The first pulsar in an eccentric binary discovered in a
GC was PSR~B2127+11C (\cite{agk+90}). This is very similar
to the original binary pulsar, B1913+16 (\cite{ht75,wnt10}),
so it does not indicate anything special is happening in
GCs. The discovery of PSR~B1802$-$07, in the GC NGC~6539
(\cite{dbl+93}) revealed a type of system unknown in the Galactic
disk: a relatively fast-spinning pulsar ($P = 23.1$\,ms) with a low-mass
($\rm M_c \sim 0.3 M_{\odot}$) companion an an eccentric
($e = 0.21$) orbit. All similar systems then known in the
Galaxy had very low ($< 10^{-3}$) eccentricities. This
indicated severe orbital perturbations by passing stars
(\cite{phi93}), something to be expected given the high
stellar densities in the cores of GCs.

The discovery of PSR~J0514$-$4002A, in NGC~1851, an MSP with
a spin period of 4.99 ms, a very eccentric ($e = 0.888$)
18.8-day orbit (\cite{fgri04}) and massive companion indicates
very conclusively that the pulsar exchanged companions
after being recycled: the companion is too massive 
($\rm M_c > 0.96 M_{\odot}$, \cite{frg07}) for
its progenitor to have recycled the pulsar to its current
spin period. This is also the case for at least another
system, PSR~J1807$-$2500B, in NGC~6544 (\cite{lfrj12}).
 Thus, this sort of system bears witness to
the very same process that lead to the recycling of so many
pulsars in GCs. Their dense environments
can produce binary pulsars (and binary systems in general) that
are truly unlike anything that binary stellar evolution can
produce in the Galactic disk. These are the systems we
designate here as ``exotic".

A total of 19 eccentric ($e > 0.2$) systems have been
discovered in GCs (e.g.~\cite{rsb+04,pdc+05,frb+08,drr11,lfrj12}),
including 7 such systems in Terzan 5 alone (\cite{rhs+05}).
The eccentricities allow at least the measurement of
the rate of advance of periastron. When this effect is due
to the effects of general relativity alone it yields
an estimate of the total mass of the binary
(\cite{dbl+93,fck+03,rhs+05,frg07,frb+08,fwbh08}). When
more Post-Keplerian measurements become available, we
can measure individual masses precisely and in some
cases test general relativity (\cite{lfrj12,jcj+06}).

\subsection{Eclipsing binaries}

Until recently, it appeared that one of the distinctive
characteristics of the pulsar population in GCs was
the large number of eclipsing binaries: 21 known at
present. Until 2009, only two were known in the Galactic
disk, the original ``Black Widow" system, PSR~B1957+20
(\cite{fst88}) and PSR~J2051$-$0827 (\cite{sbl+96}).
Furthermore, several of the eclipsing systems discovered
in GCs [e.g., PSR~B1718$-$19 in NGC~6342 (\cite{lbhb93}),
PSR~J1748$-$2446A, P and ad in Terzan 5
(\cite{ljm+90,rhs+05,hrs+06}), J0024$-$7204W
in 47~Tuc (\cite{clf+00,egc+02}), J1740$-$5340 in
NGC~6397 (\cite{dlm+01,dpm+01,fpds01}), J1701$-$3006B
in M62 (\cite{pdm+03}) and PSR~J2140$-$2310A in M30
(\cite{rsb+04})] had in some cases very extensive
eclipses and non-degenerate
companions with a few tenths of a solar mass. Because
these systems (nicknamed ``Redbacks" by Mallory Roberts)
had no counterpart in the Galaxy, they
were thought to be ``exotic", i.e., results of
exchange interactions where a radio pulsar acquires
a new main-sequence companion.

However, it was suggested that, due to its similarity with
SAX~J1808.4$-$3658 (the first accreting MSP,
\cite{wk98}), 47~Tuc~W might also represent a transitional 
object (\cite{bgb05}), not an exotic system.
The building of the LMXB-MSP bridge continued with the
discovery of PSR~J1023+0038 (\cite{asr+09}), the first
Galactic ``Redback", which showed these objects are not
restricted to GCs.
The number of eclipsing systems in the Galaxy has
since increased dramatically, in great part due to
 the launch of the Fermi satellite. The tale is told by Mallory Roberts
in these proceedings. The
``Black Widow" and ``Redback" systems are much more abundant
in the Galactic disk than previously thought. Despite this,
some eclipsing systems in GCs are likely to be truly ``exotic",
like PSR~B1718$-$19 (\cite{kkk+00}).

\subsection{Super-energetic MSPs: in GCs and in the Galaxy?}

In GCs, there are two MSPs that appear to have
unusually high magnetic fields, very large spin-down energies
and ages smaller than 30 Myr --- about
0.3 \% of the age of the clusters that host them.
One of them is the first GC pulsar, PSR~B1821$-$24.
Soon after its period derivative ($\dot{P}$) was measured
(\cite{fbtg88}), the unusual nature of the MSP was
noticed and discussed. It was deemed to be unlikely that
a contribution from the cluster acceleration could
be the cause for the anomalous $\dot{P}$.

There is firmer evidence of this for the
second ``super-energetic" MSP to be discovered,
PSR~B1820$-$30A in NGC~6624 (\cite{bbl+94}).
Initially the discoverers suggested that the
very high $\dot{P}$ was due to acceleration in
the cluster; this is a possibility given that
the cluster has a collapsed core and the pulsar is
(at least in projection) very close to the center.
However, the high $\gamma$-ray luminosity of this
object (\cite{faa+11}) implies that
this pulsar has to be energetic and quite young.

These MSPs have lifetimes $\sim 10^2$ times shorter
than the more normal MSPs and are about $10^2$ times
less abundant; therefore both types must
be forming at comparable rates. 
We list three main possibilities for their formation process:
(a) These pulsars were members of now disrupted
X-ray binaries, where spin-up went much further than
for the young, slow pulsars, but where the ``burial" of the
magnetic field was not concluded. This would nicely explain
why both objects are single,
(b) These pulsars could result from
AIC or merger-induced collapse of WDs
(\cite{ihrb08}),
(c) They might have an identical formation channel
to other MSPs (\cite{tlk12}). 
If (a) is correct, then these super-energetic MSPs
are ``exotic", i.e., only found in GCs. Otherwise
we should see similar systems in the Galaxy, likely
associated with gamma-ray sources.
As the present generation of high-resolution 20-cm
surveys probes deeper into the Galaxy for MSPs
(\cite{cfl+06,kjs10,blr+12,lbr+12})
we might soon know whether pulsars like
PSR~B1820$-$30A and
B1821$-$24A exist outside GCs or not. 
Finding them would have grand implications: it would mean
that they are forming at rates similar to normal MSPs
through the Universe and that MSPs are born with
a range of magnetic fields wider than currently
believed; the observation that the majority of MSPs
have very low B-fields would then be a selection effect
caused by low B-field MSPs being much longer lived.

\section{Is there a low-velocity NS population in the Galaxy?}

If e-capture SNe are forming NSs in GCs, they
should also be forming NSs in the Galaxy:
nothing about them is exclusive to GCs.
A low-velocity NS population has indeed been
suggested several times from proper motion data
of radio pulsars (e.g., \cite{acc02}),
but this signature is not clear and could be
due instead to projection effects (\cite{hllk05}).
This might be explained if e-capture SNe
result only from AIC of a massive O-Ne-Mg
WD (or, alternatively, from the merger of two WDs),
i.e., if it requires evolution in a binary system
(\cite{plp+04}). If this is true, then binary
systems are the place to look for evidence of NSs
formed in e-capture SNe in the Galaxy.

The binaries seem to agree. Pfahl et al.
(2002)\nocite{prps02} identified a class of high-mass,
long orbital period ($P_b >$\,30 days)
low-eccentricity X-ray binaries
(the prime example being X~Per/4U 0352+309),
in which the NSs must have been born with a
low kick velocity. Later, remarking the low eccentricities
of several double neutron star
systems, van den Heuvel (2004) suggested that the
second-born NSs in those systems were produced by 
e-capture SNe. An additional piece of evidence in these
systems is the NS mass measurements, some of them
as low as 1.25~M$_{\odot}$ (e.g., \cite{ksm+06}), as expected
from the gravitational collapse of a contracting O-Ne-Mg core
that is just beyond its Chandrasekhar mass (1.38~M$_{\odot}$,
see \cite{spr10} and references therein).

There is also some evidence for systems where the
first formed NS has a very low mass and low velocity,
(PSR~J1802$-$2124; \cite{fsk+10}), or a very small
vertical velocity and possibly a small NS mass
(PSR~J1949+3106; \cite{dfc+10}), however the
massive WD progenitors 
should have diminished post-SN velocities of these binaries.
Another case is PSR~J1903+0327
(\cite{crl+08}), the first MSP discovered in the ALFA
pulsar survey (\cite{cfl+06}). This anomalous pulsar
was likely formed in a triple system (\cite{fbw+11,phln11}),
but the preservation of the triple requires
a very small kick velocity (\cite{pcp12}); this is consistent
with the very small peculiar velocity of this system (\cite{fbw+11}).

Systems where the first NS formed in e-capture SNe should
constitute a dynamically separate NS population in the Galaxy
with very low scale height; furthermore many of the NSs
should have distinctively low masses. Proper motion measurements
of the many MSPs being discovered in the current
high-resolution 20 cm surveys should be able to determine
whether such a population exists and determine its size.

%============================================================

\section*{Acknowledgments}

I thank the financial support by the European Research Council
for the ERC Starting Grant BEACON under contract no. 279702.

%==============================================================================%

\end{document}